\documentclass[reprint,aps,pra]{revtex4-1}

\usepackage{graphicx}
\usepackage{dcolumn}
\usepackage{bm}
\usepackage{amsmath}
\usepackage{amsthm}
\usepackage{amssymb}
\usepackage{float}
\usepackage{hyperref}
\usepackage{color}
\usepackage{epstopdf}
\usepackage{cleveref}
\usepackage[svgnames]{xcolor}
\hypersetup{hidelinks,colorlinks=true,allcolors=DarkBlue}

\begin{document}

\preprint{APS/123-QED}

\title{Structural monitoring system with fiber Bragg grating sensors: \\
Implementation and software solution}
\author{Aleksey Fedorov$^{1,2}$}\email{akf@rqc.ru}
\author{Vladimir Lazarev$^{2}$}\email{sintetaza@mail.ru}
\author{Ilya Makhrov$^{2}$}
\author{Nikolay Pozhar$^{2}$}
\author{Maxim Anufriev$^{2}$}
\author{Alexey Pnev$^{2}$}
\author{Valeriy Karasik$^{2}$}
\affiliation
{
\mbox{$^{1}$Russian Quantum Center, Skolkovo, Moscow 143025, Russia}
\mbox{$^{2}$Bauman Moscow State Technical University, Moscow 105005, Russia}
}

\date{\today}

\begin{abstract}
We present a structural health monitoring system for nondestructive testing of composite materials based on the fiber Bragg grating sensors and specialized software solution.
The developed structural monitoring system has potential applications for preliminary tests of novel composite materials as well as real-time structural health monitoring of industrial objects.
The software solution realizes control for the system, data processing, and alert of an operator.
\end{abstract}
                              
\maketitle

\section{Introduction}

The implementation of nondestructive structural health monitoring (SHM) systems for industrial objects is a challenging problem. 
In general, SHM is a method of integrity analysis of an object via real-time measurements of a class of parameters, which characterize the object's integrity. 
The main assignment of SHM systems is an alert of the system operator in case of exceeding the measured parameters of a certain critical value, where this value is determined from preliminary tests. 
Problems of a design of SHM are of current interest for civil and industrial buildings, pipelines, airplanes, water transport, and in many other scopes \cite{review,Tennyson,Tanaka}.

Composite materials have unique physical and chemical properties \cite{composite}.
These materials have corrosion resistance, high heat resistance, wear resistance, rigidity and low mass. 
The properties make them useful in a wide spectrum of technological applications, {\it e.g.}, in space technologies, aviation engineering, shipbuilding, and etc.
The design of SHM systems for composite materials is especially topical \cite{review}. 

There are a two main approaches to design of a measurement part of SHM systems.
The first method of SHM involves mounting sensors directly into the material.
The second one is monitoring of the construction by the installation of sensors to certain elements of the structure. 
In the case of composite material structures, SHM systems are mostly based on embedding the sensors into the material \cite{review}. 

In this way, the main question is what type of sensors is the most useful for SHM of composite materials.
Practically, fiber optic sensors form the most useful class of sensors for SHM systems.
There are several types of fiber-based optic sensors.
The most common solutions are
sensors based on measuring light intensity using changes in fiber curvature or reflected light from a mirrored surface,
the Fabry--Perot sensors, 
and the fiber Bragg grating (FBG) sensors.
Using of terahertz vision system for nondestructive testing of composite materials is of current interest as well \cite{Stoik, Yakovlev}. 

Using of an embedded FBG sensors for measurement of distributed stress and strain fields of the composite material has significant advantages \cite{review, Othonos,Vasilev}. 
First of all, FBG sensors can be easily embedded into materials to provide damage detection or internal strain field mapping. 
The FBG sensors supposed to be located in fiber optic cable mounted in the sample. 
Furthermore, the key feature of FBGs sensors is that the information about perturbations is encoded in wavelength. 
Indeed, the main principle of FBG sensors work is transform of spectral or phase characteristics of the test optical signal propagating in the sample according to the perturbations of a measured physical quantity. 
Due to their high noise resistance to the effects of non-informative influencing factors \cite{Lazarev1, Lazarev2}, FBG sensors attract increased attention \cite{review}. 
FBG is a periodic structure of refractive index which is formed in a fiber core (see, Fig. 1). 
SHM systems have a measurement part, which typically consists from: FBG sensors, a source of test optical signal, and a measuring device for the signal reflected from FBG.
Important problem is the embedding the sensor into the material sample. 
Useful solution of this problem is proposed in \cite{Lazarev1,Lazarev2,Lazarev3,Lazarev4}. 

\begin{figure}
\begin{center}
\includegraphics[width=0.25\textwidth]{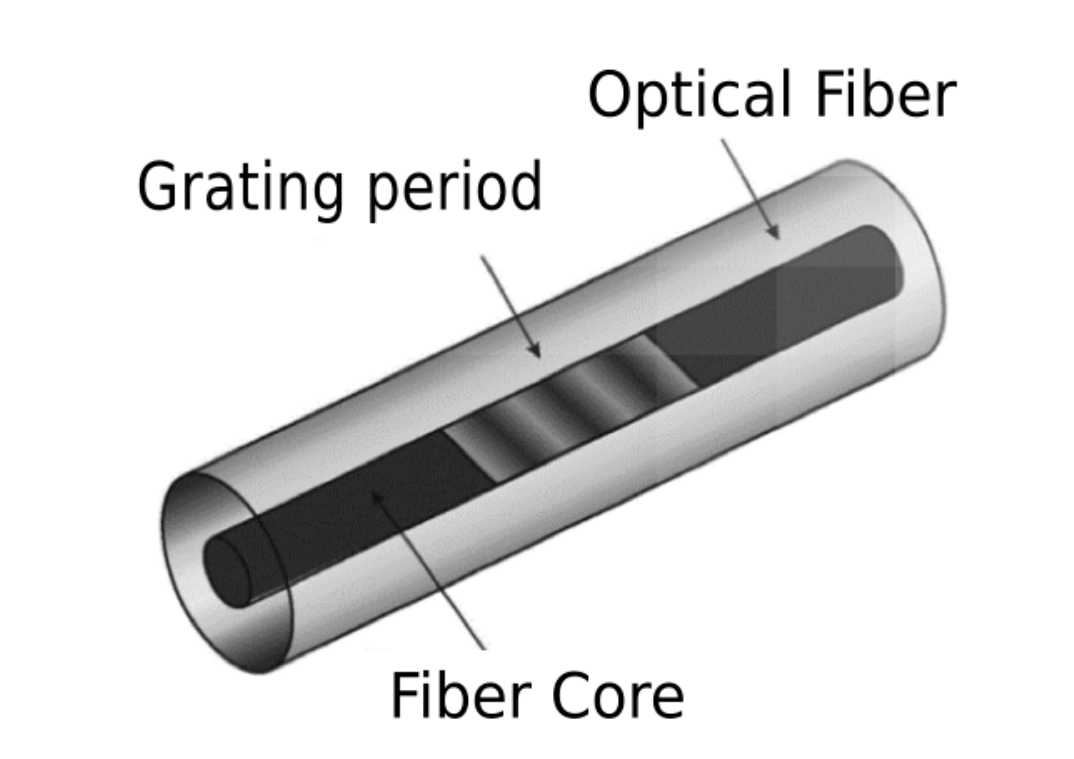}
\caption{Structure of FBG sensor. Typical size of sensors is $5-7$ mm with typical grating period being $500$ nm.}
\end{center}
\end{figure}

\begin{figure*}[t]
\begin{centering}
\includegraphics[width=0.45\textwidth]{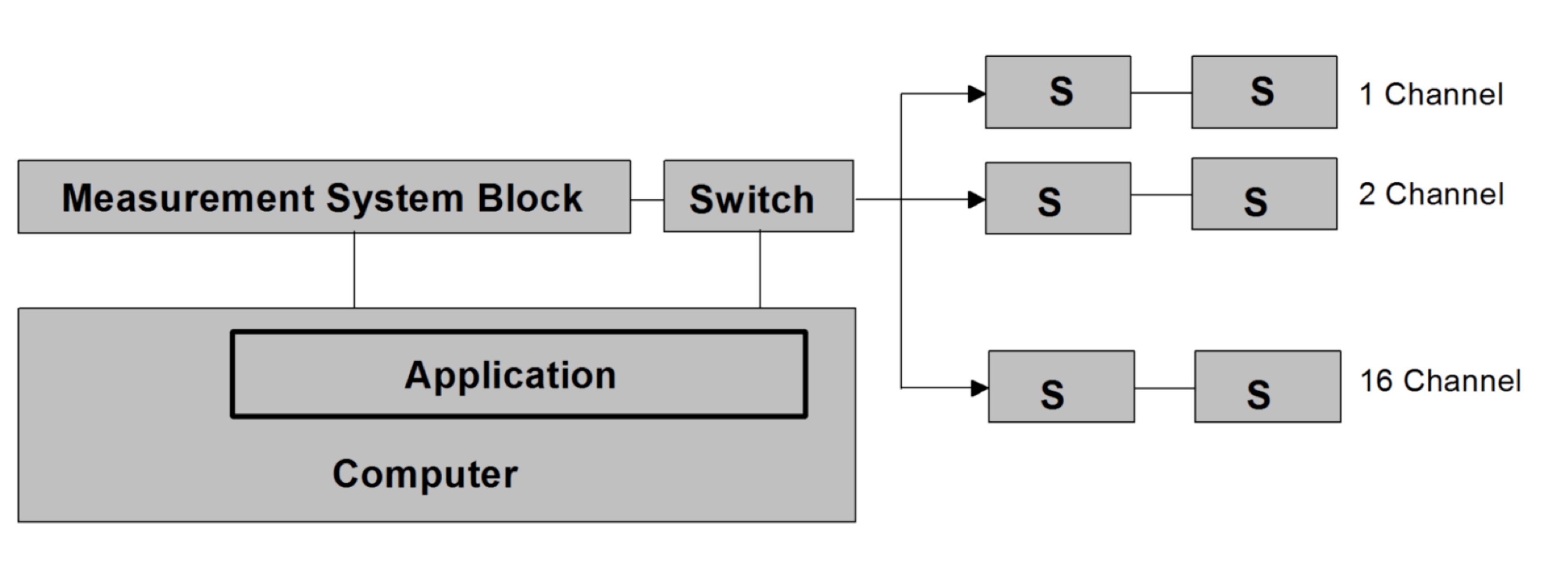}
\includegraphics[width=0.45\textwidth]{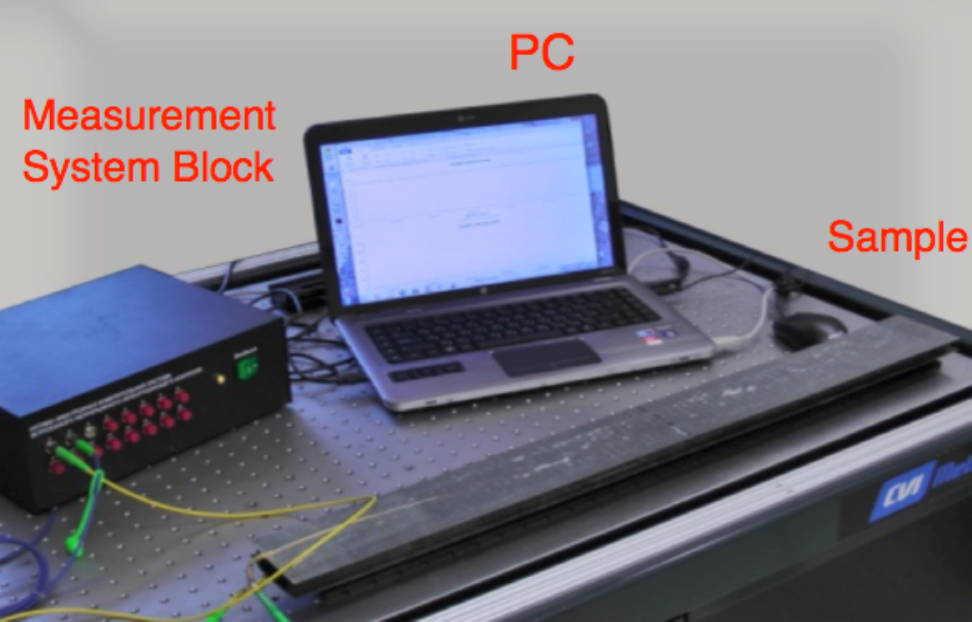}
\end{centering}
\caption
{
{\it Left}: Structural monitoring system scheme. 
Channel of the measurement system is a fiber optic cable. 
There are 16 channels embedded in the sample. 
Each channel includes up to 15 sequentially arranged Bragg sensors (S). 
Each sensor has a central wavelength of the reflected signal. 
{\it Right}: Implementation of SHM system.
}
\label{fig:1}
\end{figure*}

On the other hand, practical application and implementation of SHM systems require using of a software solution for control, fast collection, reliable storage, and effective processing of large data flow from the measurement part in real time.
Indeed, the data proceeding is an another kind of problem, which can be solved efficiently only using software solution. 
The software part is needed for the solution of the following classes of problems:
processing of the measurements from FBG sensors,
visualization of the measurements results, and
``user--measurement part'' interaction, and control for the system.

In the present work, we present developed SHM system for composite materials based on the fiber Bragg grating sensors. 
We present the software solution for developed SHM system. 
The system has two main scopes of applications. 
The first is preliminary tests system for novel composite materials.
The other one is application for real-time SHM with an alert for an operator.

\section{Implementation of SHM system}

There are two physical reasons for the fluctuation of the FBG sensor central wavelength $\Delta\lambda$: the mechanical stress $\Delta{l}$ and the temperature $\Delta{T}$ variations \cite{Vasilev}. 
The central wavelength of Bragg grating as function of the strain and the temperature can be presented as follow 
\begin{equation}\label{equation}
	\Delta\lambda=
	2\left(\Lambda\frac{\partial{n}}{\partial{T}}+n\frac{\partial{\Lambda}}{\partial{T}}\right)\Delta{T}+
	2\left(\Lambda\frac{\partial{n}}{\partial{l}}+n\frac{\partial{\Lambda}}{\partial{l}}\right)\Delta{l},
\end{equation}
where $n$ is the effective refractive index and $\Lambda$ is the grating period.

The first term of the right-hand side of Eq. (\ref{equation}) describes the effect of temperature on the wavelength 
\begin{equation}
	\Delta\lambda=\lambda(a_{\Lambda}+a_n)\Delta{T},
\end{equation}
where $a_{\Lambda}$ is the thermal expansion coefficient and $a_n$ is the thermo-optic coefficient of the fiber. 

In turn, second part of the right-hand side of Eq. (\ref{equation}) describes the nature of the dependence of the central wavelength $\lambda$ from the applied strain, which can be represented in the form
\begin{equation}
	\Delta\lambda=\lambda(1-p_e)\varepsilon, \qquad p_e=\frac{n^2}{2}\left[p_{12}-\nu(p_{11}+p_{12})\right],
\end{equation}
where $\varepsilon$ is the relative strain, $p_e$ is strain constant for optical fiber, $p_{11}$ and $p_{12}$ are the Pockels coefficients in the optical stress tensor, and 
$\nu$ is the Poisson coefficient. 

Functioning of the measurement part of SHM system with FBG sensors is based on the following principles (Fig. 2).
Each channel of the measurement system is a fiber optic cable. 
There are 16 channels, embedded in the composite material. 
All of these sensors are sequentially arranged Bragg sensors. 
Sensor have a unique central wavelength of the reflected signal in channel.

The aim of SHM systems is to measure the strain impact, but for the reliability of the data it is necessary to consider the effect of temperature. 
In other words, the idea is to isolate temperature and strain effects. 
For this reason, strain-invariant (called ``temperature'') sensors in FBG are used for measuring the temperature effect only. 

The measurement part (measurement system block, Fig. 1) is intended for interrogating the sensors by applying the test input of each channel of broadband optical signal (the bandwidth is $\Delta{\lambda_{\rm test}}\approx{40}$ nm). 
The part makes the measurement of wavelengths reflected from the Bragg grating, {\it i.e.}, it works as an optical spectrum analyzer.
Signal reflected from FBG sensors is the input to an optical switch (switch, Fig. 1), which is recorded after the measuring system. 
The switch produces a serial connection channel to measurement system block, which performs  measurements of the central wavelengths of the reflected radiation from each FBG sensor (see Fig 2).
These measurement system block and optical switch are connected to PC via the standard USB-interface.

\begin{figure*}[t]
\begin{centering}
\includegraphics[width=0.68\textwidth]{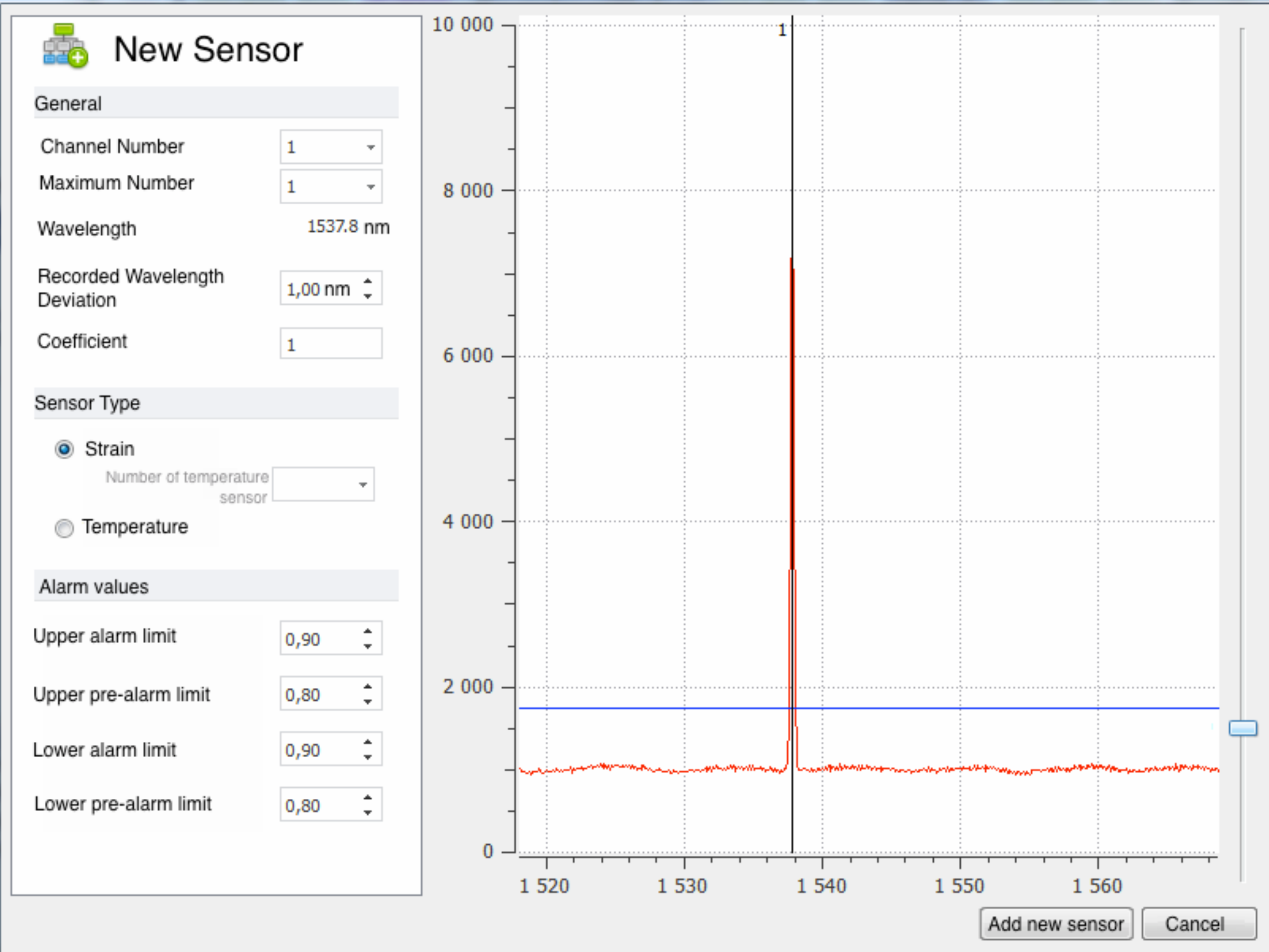}
\end{centering}
\vskip -4mm
\caption
{
Software interface: adding of a new sensor in the measurement part of SHM system.
}
\end{figure*}

Start our consideration from the simplest case of a one-channel measurement part. 
Suppose that there are $n$ sensors in the channel, where $n-k$ is a quantity of Bragg sensors and $k$ is a quantity of temperature Bragg sensors. 
Initial data for the structural monitoring system is the set of  the reflection wavelength
\begin{equation}
	\lambda=(\lambda_1,\dots,\lambda_{n}),
\end{equation}
where $\lambda_i$ is the reflection wavelength of $i$th sensor. 
The result of the measurement in the channel is following set 
\begin{equation}
 	\lambda^{r}=(\lambda_1^{r},\dots,\lambda_{n}^{r}),
\end{equation}
where $\lambda_i^{r}$ is the measured reflecting wavelength of $i$th sensor. 

First, initial reflection wavelength of strain-invariant sensors compared with measured for identifying a temperature effect $\delta{T}$. 
The set of wavelengths  
\begin{equation}
	{\lambda^{r}}\mapsto{\bar{\lambda}^{r}}=(\bar{\lambda}_1^{r},\dots,\bar{\lambda}_{n}^{r})
\end{equation}
is formed from $\lambda^{r}$ for excluding the temperature effect. 
Second, $i$th component of the vector ${\bar{\lambda}^{r}}$ compared with $i$th component of the vector $\lambda$ for temperature sensors. 
Main aim of this procedure is to identify sensors with non-zero difference 
\begin{equation}
	\delta_i=\bar{\lambda}^{r}-\lambda_i.
\end{equation}
Finally, from grating period $\Lambda$, number of sensor and the sample length $l$ it is possible to localize the position of strain effects of the sample. 

It is clear that for multi-channel system, {\it e.g.}, for our system, there is matrix $\lambda=\lambda_{ij}$ where $i$ is the sensor number and $j$ is the channel number. 
Matrix elements, obviously, $\lambda_{ij}$ are the reflection wavelengths of the $i$th sensor in the $n$th channel. 
However, general procedure of data processing is the same.

We describe the principles of operation established for that purpose multi-threaded software applications. 
We note that there is a energy depend memory part in the measurement system block and its DLL library used for getting data. 
The GUI software application is a system of tabs with different settings and for displaying data. 

Screen of the main control page of developed software solution is presented on the Fig. 3. 
From software point of view, every sensor is characterized via set of parameters. 
As it has been mentioned before, every sensor measures both temperature and mechanical strain. 
In this way, first parameter of the sensor is a type. 
In addition, there are channel number, lower and upper limit among them. 

On practice, special sensors which are sensitive only to changes in temperature are used. 
This allows to correct measurement results obtained from other sensors.
After the addition of sensors in the channels formed by the summary table, which allows the operator to control the number and types of currently active sensors. 
The operator at any time is able to move the sensors for tracking in real time. 

\begin{figure*}[t]
\begin{centering}
\includegraphics[width=0.85\textwidth]{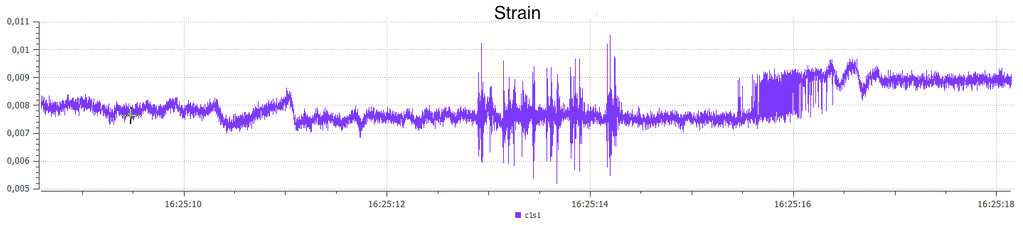}
\includegraphics[width=0.85\textwidth]{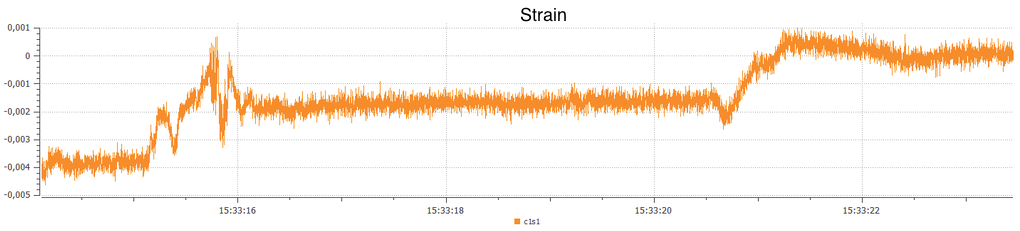}
\end{centering}
\caption{The results of testing of composite materials: strain as function of time.}
\end{figure*}

Via DLL libraries optical switch and measurement system implemented algorithms for channel management in the optical switch and selection the optimum mode of data collection, 
{\it i.e.}, a combination of channels, the number of active sensors in the channel sample rate sensors. 

Additionally, the software part solves the problem of data visualization. 
Fig. 2 presents our laboratory experimental setup in accordance with the general experimental setup (Fig. 2).
Fig. 4 shows the results of processing measurement signals from the measurement equipment  in the test laboratory strain of the composite material in various load conditions: this is strain as function of time. 
It should be noted that an important element in developing the software module is a data storage as an integral part of the system when used in practice, is the storage and archiving.

\section{Conclusion} 

Structural monitoring systems are actively used to solve problems of the analysis of the dynamics of aging infrastructure in order to prevent their destruction. 
Developed systems allows to realize non-destructive structural health monitoring of structures made of composite materials. 
Software allows to perform the following kinds of problem: control the measurement part (activation of sensors, one of their number, channel selection and polling frequency sensors).
Visualization of measurement results in real time is obtained via using a software application. 
The use of software and hardware optoelectronic measurement system integrated continuous non-destructive testing is promising for a wide range of engineering problems. 
The software part of the complex can be easily modified to the case of using another type of sensor or monitoring by the sensors to certain elements of the structure.

{\bf Acknowledgments}.
We thank Scientific-Educational Center ``Photonics and Infrared Technology'' for support and K.I. Zaytsev for useful comments.
AKF is a Dynasty Foundation Fellow.
We are grateful to the organizers of the 23rd Laser Physics Workshop (Sofia, July 14--18, 2014) for kind hospitality.

\end{document}